\documentclass[11pt]{article}
\usepackage{mystyle1}
\usepackage{color, times}
\usepackage{epsfig}

\let\eps\varepsilon
\newcommand{\reals}{\mathbb{R}}

\def\MST{|\mbox{MST}|}
\def\diam{\mbox{diam}}

\newcommand{\denselist}{\itemsep -2pt\parsep=-2pt\partopsep -2pt}
\newcommand{\bitem}{\begin{itemize}\denselist}
\newcommand{\eitem}{\end{itemize}}
\newcommand{\benum}{\begin{enumerate}\denselist}
\newcommand{\eenum}{\end{enumerate}}

\newcommand{\myparagraph}{\vspace*{-4mm}\paragraph}

\title{The Emergence of Sparse Spanners and \\Greedy Well-Separated Pair Decomposition}
\author{Jie Gao\thanks{%
Department of Computer Science, Stony Brook University, Stony Brook,
NY 11794, USA, {\tt \{jgao, dpzhou\}@cs.sunysb.edu}.} \and Dengpan
Zhou$^\ast$}
\begin{document}
\maketitle


\begin{abstract}
A spanner graph on a set of points in $\reals^d$ contains a shortest
path between any pair of points with length at most a constant
factor of their Euclidean distance. A spanner with a sparse set of
edges is thus a good candidate for network backbones, as desired in
many practical scenarios such as the transportation network and
peer-to-peer network overlays. In this paper we investigate new
models and aim to interpret why good spanners `emerge' in reality,
when they are clearly built in pieces by agents with their own
interests and the construction is not coordinated. Our main
result is to show that the following algorithm generates a
$(1+\eps)$-spanner with a linear number of edges, constant average
degree, and the total edge length as a small logarithmic factor of the
cost of the minimum spanning tree. In our algorithm, the points
build edges at an \emph{arbitrary} order. When a point $p$ checks on
whether the edge to a point $q$ should be built, it will build this
edge only if there is no existing edge $p'q'$ with $p'$ and $q'$ at
distances no more than $\frac{1}{4(1+1/\eps)} \cdot |p'q'|$ from $p,
q$ respectively. Eventually when all points have finished checking
edges to all other points, the resulted collection of edges forms a
sparse spanner as desired. This new spanner construction algorithm can be extended to a metric space with constant doubling dimension and admits a local routing scheme to find the short paths.

As a side product, we show a greedy algorithm for constructing
linear-size well-separated pair decompositions that may be of interest
on its own. A well-separated pair decomposition is a collection of
subset pairs such that each pair of point sets is fairly far away
from each other compared with their diameters and that every pair of
points is `covered' by at least one well-separated pair. Our greedy
algorithm selects an \emph{arbitrary} pair of points that have not yet
been covered and puts a `dumb-bell' around the pair as the well-separated
pair, repeats this until all pairs of points are covered. When the algorithm finishes, we show only a linear number of pairs is generated, which is
asymptotically optimal.
\end{abstract}

\section{Introduction}

A geometric graph $G$ defined on a set of points $\P\subseteq
\reals^d$ with all edges as straight line segments of weight equal
to the length is called a {\em Euclidean spanner}, if for any two
points $p, q\in P$ the shortest path between them in $G$ has length
at most $\lambda \cdot |pq|$ where $|pq|$ is the Euclidean distance. The
factor $\lambda$ is called the \emph{stretch factor} of $G$ and the graph
$G$ is called a $\lambda$-spanner. Spanners with a sparse set of edges
provide good approximations for the pairwise Euclidean distances and
are good candidates for network backbones. Thus, there has been a
lot of work on the construction of Euclidean spanners in both the centralized
setting~\cite{e-sts-00,geometric07narasimhan} and the distributed
setting~\cite{peleg00}.

In this paper we are interested in the emergence of good Euclidean
spanners formed by uncoordinated agents. Many real-world
networks, such as the transportation network, the Internet backbone
network, the flight network, are good spanners --- one can typically
drive from any city to any other city in the U.S. with the total
travel distance at most a small constant times their straight line
distance; similarly, one can find connecting flights between any two
towns with the total flight distance not much more than the direct
flight distance, possibly by flights of different airlines. The same
thing happens with the Internet backbone graph as well. However,
these large networks are not owned or built by any single authority.
They are often assembled with pieces built by different governments
(federal, states, or county governments), different airline
companies, or different ISPs. These agents have their own agenda and
interests. It does not seem to be the case that they have carefully
planned the construction in collaboration before hand and clearly
the edges are not built all at the same time.
Nevertheless altogether they provide a convenient sparse spanner for
the users that is close to the best spanner one can find in a
centralized coordinated setting. The work in this paper is largely
motivated by this observation and we would like to interpret and understand
why a good Euclidean spanner is able to `emerge' from these agents
incrementally.

From the application's perspective, we are interested
in the construction of nice network overlay or infrastructure
topologies used in many distributed network services such as
transportation network, peer-to-peer (P2P) file sharing and content
distribution applications~\cite{lua05survey}. Such overlay or
infrastructure networks are constructed in a distributed manner
without centralized control, to achieve robustness and scalability.
The agents self-organize themselves in a overlay network topology by
choosing other agents to connect to directly. While prior work in
overlay design has focused on system robustness and network
scalability, a significant amount of work in recent years have
focused on reducing routing
delay~\cite{chu01enabling,ratnasamy02topologically,kwon02topology,wang05network,zhang08distributed}.
Most of these work adapt the current overlay topology to respect the
topology of the underlying network and only show by experiments the
improvement of the average delay. The fundamental question still yet
to be answered is as follows~\cite{ratnasamy02topologically}: for a
peer in the P2P network, what set of neighbors should it connect to,
such that the shortest path routing latency on the resultant overlay
is low, compared with the minimum delay in the underlying network?
Obviously a spanner graph would be a good solution for the overlay
construction, yet there is no centralized authority in the P2P
network that supervises the spanner construction and the peers may
join or leave the network frequently (the system churn rate is
high). The work in this paper initiates the study of the emergence
of good spanners in the setting when there is little coordination
between the agents and the users only need a modest amount of
incomplete information of the current overlay topology.

\myparagraph{Our contribution.} We consider in this paper the
following model that abstracts the scenarios explained earlier.
There are $n$ points in the plane. Each point represents a separate
agent and may build edges from itself to some other points by the
strategy to be explained later. The edges in the final graph is the
collection of edges built by all the agents. The agents may decide
to build edges at different point in time. When an agent $p$ plans
on whether an edge from itself to another point $q$ should be built
or not, $p$ checks to see whether there is already an edge from some
points $p'$ to $q'$ such that $|pp'|$ and $|qq'|$ are both within
$\frac{1}{4(1+1/\eps)} \cdot |p'q'|$ from $p$ and $q$ respectively.
If not, the edge $pq$ is built, otherwise it is not. This strategy
is very intuitive --- if there is already a cross-country highway
from Washington D.C. to San Francisco, it does not make economical
sense to build a highway from New York to Los Angeles. We assume
that each agent will eventually check on each possible edge from
itself to all the other points, but the order on who checks which
edge can be \emph{completely arbitrary}. With this strategy, the
agents only make decisions with limited information and no agent has
full control over how and what graph will be constructed. It is not
obvious that this strategy will end up with a sparse spanner on all
points. It is even not clear that the graph will be connected.

The main result in this paper is to prove that with the above
strategy executed in \emph{any} arbitrary order, the graph built at
the end of the process is a sparse spanner graph with the following
properties: \bitem
\item Between any two points $p, q$, there is a path with stretch $1+\eps$ and $O(|pq|^{1/(1+2/\eps)})$ hops.
\item The number of edges is $O(n)$.
\item The total edge length of the spanner is $O(\MST\cdot \log{\alpha})$, where $\alpha$ is the \emph{aspect ratio}, i.e., the ratio of the distance between the furthest pair and the closest pair, and $\MST$ is the total edge length of the minimum spanning tree of the point set. Clearly $\MST$ is a lower bound on the total edge length of any constant stretch spanner.
\item The degree of each point is $O(\log\alpha)$ in the worst case and $O(1)$ on average.
\eitem An example of such a spanner is shown in
Figure~\ref{fig:spanner} in the Appendix. The above results for Euclidean space can be extended to a metric with constant doubling dimension. We also show that
the spanner can be constructed for $n$ agents with the help of a near neighbor oracle such that \bitem
\item Only a total number of $O(n\log \alpha)$ messages need to be exchanged between the agents during the entire construction process.
\item The spanner topology is implicitly stored on the nodes with each node's storage cost bounded by $O(\log\alpha)$. It turns out with this representation the nearest neighbor of each node is included in the information stored on each node.
\item And yet simply with the partial information stored at each node, there is a local distributed algorithm that finds a $(1+\eps)$-stretch path with maximum $O(|pq|^{1/(1+2/\eps)})$ hops between any two nodes.
\eitem

To explain how this result is proved, we first obtain as a side
product the following \emph{greedy} algorithm for computing a
well-separated pair decomposition of the points with optimal size. A
pair of two sets of points, $(A, B)$ with $A, B$ as subsets of $\P$,
is called \emph{$s$-well-separated} if the smallest distance between
any two points in $A, B$ respectively is at least $s$ times greater
than the diameters of $A$ and $B$. An \emph{$s$-well-separated pair
decomposition} ($s$-WSPD for short) is a collection of
$s$-well-separated pairs $\W=\{(A_i, B_i)\}$ such that for any pair
of points $p, q\in \P$ there is a pair $(A, B)\in \W$ with $p\in A$
and $q\in B$. The size of an $s$-WSPD is the number of point set
pairs in $\W$. Well-separated pair decomposition (WSPD) was first
introduced by Callahan and Kosaraju~\cite{ck-dmpsa-95} and they
developed algorithms for computing an $s$-WSPD with linear size for
points in $\reals^d$. Since then WSPD has found many applications in
computing $k$-nearest neighbors, $n$-body potential fields,
geometric spanners and approximate minimum spanning
trees~\cite{callahan93faster,c-opann-93,ck-dmpsa-95,ck-adcpn-95,ams-rdags-94,admss-esstl-95,ns-asfeg-00,levcopoulos02improved,glns-adogg-02,e-dpssd-02}.

So far there are two algorithms for computing optimal size WSPD, one
in the original paper~\cite{ck-dmpsa-95} and one in a later
paper~\cite{gao06deformable}. Both of them use a hierarchical
organization of the points (e.g., the fair split tree
in~\cite{ck-dmpsa-95} and the discrete center hierarchy
in~\cite{gao06deformable}) and output the well-separated pairs in a
recursive way. In this paper we show the following simple
\emph{greedy} algorithm also outputs an $s$-WSPD with linear size. In
particular, we take an \emph{arbitrary} pair of points $p, q$ that
is not yet covered in any existing well-separated pair, and consider the pair
of subsets $(B_r(p), B_r(q))$ with $r=|pq|/(2s+2)$ and $B_r(p)$
($B_r(q)$) as the set of points of $\P$ within distance $r$ from $p$
($q$). Clearly $(B_r(p), B_r(q))$ is an $s$-well-separated pair and
now all the pairs of points $(p', q')$ with $p'\in B_r(p)$ and
$q'\in B_r(q)$ are covered. The algorithm continues until all pairs
of points are covered. We show that, no matter
in which order the pairs are selected, the greedy algorithm will always output a linear number of well-separated pairs.

The key idea in proving the linear size WSPD generated by the greedy
algorithm is to show that at most a constant number of the generated
well-separated pairs can be mapped to each well-separated pair
generated by the deformable spanner~\cite{gao06deformable}, a data
structure that has found many applications in proximity search with
efficient update algorithm in both the
kinetic~\cite{gao06deformable} and dynamic
settings~\cite{roditty07fully,gottlieb08improved}. It has been shown
in~\cite{gao06deformable} that the deformable spanner implies a WSPD
with linear size. Thus the greedy algorithm also finds a linear
number of pairs. The greedy WSPD also has a number of nice properties (not
necessarily carried by the WSPD constructed in~\cite{ck-dmpsa-95})
as to be shown in more details later. The greedy
WSPD algorithm may be of interest by itself.

Well-separated pair decomposition is deeply connected to geometric
spanners. In fact, any WSPD will generate a spanner graph if one
puts an edge between an arbitrary pair of points $p, q$ from each
well-separated pair $(A, B)\in
\W$~\cite{ams-rdags-94,admss-esstl-95,ns-asfeg-00,levcopoulos02improved}.
The number of edges in the spanner equals to the size of $\W$. In
the other direction, the deformable spanner
in~\cite{gao06deformable} implies a WSPD of linear size. The
connection is further witnessed in this paper --- in our spanner
emergence algorithm, each agent $p$ constructs an edge to $q$ only when
there is no nearby edge connecting points near $p$ to points near
$q$, this simple rule implies a WSPD generated in a greedy manner.
Hence our spanner construction in an uncoordinated manner inherits many nice properties of the greedy WSPD.

Last, this paper focuses on the case when the points are
distributed in the Euclidean space. The basic idea extends naturally to metrics with constant doubling
dimensions~\cite{gupta03bounded,ng02predicting,plaxton97accessing}, as the main technique involves essentially various forms
of geometric packing arguments.

\myparagraph{Related work.}

The model and the philosophy in this paper are related
to the network creation
game~\cite{fabrikant03network,corbo05price,jansen07stability,albers06susanne,moscibroda06topologies}.
Fabrikant \etal~\cite{fabrikant03network} was the first to introduce
the network creation game, in order to understand the evolution of
network topologies by selfish agents. The model used there (and in
follow-up papers) assigns a cost function to each agent that
captures the cost paid by the agents to build connections to others
minus the benefit received from the resulted network topology. The
agents play a game by minimizing their individual costs. Almost all
these papers use a unit cost for each edge and they deviate in how
the benefit of `being connected to others' is modeled. These papers
are interested in the existence of Nash equilibria and the price of
anarchy of Nash equilibria. There are two major open questions along
this direction. First, the choice of cost functions is heuristic ---
often some intuitive cost functions are selected. There is little understanding on what cost function best
captures the reality yet small variation in the cost function may
result in big changes in the network topologies at Nash equilibria.
It is also not easy to execute the game in practice --- either
because the selfish agents may face deadlines and have to decide on
building an edge or not immediately and sometimes the edges already
built cannot be removed later (e.g., in the development of the
transportation network by different local governments), or because
the agents do not have the big picture and the current strategies of
all other agents (e.g., in the P2P setting). The second problem is
that there is not much understanding of the topologies at Nash
equilibria. Some of the topologies at Nash equilibria in these
papers are very simplistic topologies such as trees or complete
graphs (and these topologies do not show up often in real world).
For other more sophisticated topologies, there is not much
understanding of their characteristics and therefore it is not clear
whether these topologies are desirable. Our model is connected to
the game theoretic model in the way that we also try to relax the
requirement of a centralized authority in the graph construction,
yet we also incorporate practical considerations that may not allow the
agents from playing games to reach a Nash equilibrium. We believe
such models and good algorithms under these models worth further
exploration and this paper makes a first step along this line.

In the vast amount of prior literature on geometric spanners, there
are three main ideas: $\Theta$-graphs, the greedy spanners, and the
WSPD-induced spanners~\cite{geometric07narasimhan}. We will review
two spanner construction ideas that are most related to our
approach. The first idea is the path-greedy spanner
construction~\cite{cdns-nsrgs-95,dhn-oss3d-93,dn-facse-97,dns-nwwme-95}.
All pairwise edges are ordered with non-decreasing lengths and
checked in that order. An edge is included in the spanner if the
shortest path in the current graph is longer than $\lambda$ times
the Euclidean distance, and is discarded otherwise. Variants of this
idea generate spanners with constant degree and total weight
$O(\MST)$. This idea cannot be applied in our setting as edges
constructed in practice may not be in non-decreasing order of their
lengths and in a P2P network with high churn rate it is too much
overhead to compute the shortest path length in the current overlay
network (while checking the distance between two nodes, i.e., the
path length in the underlying network topology, can be done easily
with a \textsc{TraceRoute} command). The second idea is to use the
gap property~\cite{cdns-nsrgs-95} --- the sources and sinks of any
two edges in an edge set are separated by a distance at least
proportional to the length of the shorter of the two edges and their
directions are differed no more than a given angle. The gap-greedy
algorithm~\cite{as-ecbds-97} considers pairs of points, again, in
order of non-decreasing distances, and includes an edge in the
spanner if and only if it does not violate the gap property. The
spanner generated this way has constant degree and total weight
$O(\MST)$. Compared with our algorithm, our strategy is a relaxation
of the gap property in the way that the edges in our spanner may
have one of their endpoints arbitrarily close (or at the same
points) and we have no restriction on the direction of the edges.
The proof techniques are
also quite different. The proof for the gap greedy algorithm
requires heavily plane geometry tools and our proof technique only
uses packing argument and can be extended to the general metric
setting as long as a similar packing argument holds. To get these
benefit our algorithm has slightly worse upper bounds on the spanner
weight by a logarithmic factor.

Spanner construction for metric space of constant doubling dimension has been proposed before~\cite{chan05hierarchical,small06chan,Har-Peled05fast}. These algorithms are centralized.

\myparagraph{Organization.} In the rest of the paper we first elaborate the spanner construction in an uncoordinated manner and then show the connection of the spanner with the greedy WSPD. We then show the nice properties of both the greedy WSPD and our spanner. At the end, we describe how to apply the spanner in a decentralized setting to support low-storage spanner representation and efficient local low-stretch routing.

\section{Uncoordinated spanner construction and a greedy algorithm for WSPD}\label{sec:spanner}

Assuming $n$ points in $\reals^d$, each point represents an
agent. We consider the following algorithm for constructing a sparse
spanner with stretch factor $s$ in an uncoordinated way. For any
point $p$, denote by $B_r(p)$ the collection of points that are
within distance $r$ from point $p$, i.e., inside the ball with
radius $r$ centered at $p$.

\myparagraph{Uncoordinated spanner construction.} Each point/agent $p$ will check to see whether an edge from itself to another point $q$ should be constructed or not. At this point there might be some edges already constructed by other agents. The order of which agent checks on which edge is completely \emph{arbitrary}. Specifically, $p$ performs the following operation:

\emph{Check where there is already an edge $p'q'$ such that $p$ and
$q$ are within distance $\frac{|p'q'|}{2(s+1)}$ from $p', q'$
respectively. If so, $p$ does not build the edge to $q$. Otherwise,
$p$ will build an edge to $q$.}

This incremental construction of edges is executed by different
agents in a completely uncoordinated manner. We assume that no two
agents perform the above strategy at exactly the same time. Thus
when any agent conducts the above process, the decision is based on
the current network already constructed. The algorithm terminates
when all agents finish checking the edges from themselves to all
other points. In this paper we first study the properties of
the constructed graph $G$ by these uncoordinated behaviors. We will discuss later in Section~\ref{sec:spanner-construction} a proper complexity model
for the uncoordinated construction in a distributed environment and
also bound the computing cost of this spanner.

Before we proceed with our proof, we first realize the following
invariant is maintained by the graph $G$. The proof follows
immediately from the construction of $G$.
\begin{lemma}\label{lem:separation-lemma}
\benum
\item For any edge $pq$ that is not in $G$, there is another edge $p'q'$ in $G$ such that $|pp'|\leq |p'q'|/(2s+2)$, $|qq'|\leq |p'q'|/(2s+2)$.
\item For any two edges $pq$, $p'q'$ in the constructed graph $G$, suppose that $pq$ is built before $p'q'$, then one of the following is true: $|pp'| > |pq|/(2s+2)$ or $|qq'|> |pq|/(2s+2)$.
\eenum
\end{lemma}

\commented{A slight improvement of the algorithm is to check only the points
within distance $|pq|/s$ from $p$ and $q$ for the possible edge
$p'q'$. Essentially, if an edge $p'q'$ exists such that $|pp'|\leq
|p'q'|/(2s+2)$, $|qq'|\leq |p'q'|/(2s+2)$, we know by triangular
inequality that:
\[
\begin{array}{ll}
|pp'|\leq |p'q'|/(2s+2) \leq (|pq|+|pp'|+|qq'|)/(2s+2)\,;\\
|qq'|\leq |p'q'|/(2s+2) \leq (|pq|+|pp'|+|qq'|)/(2s+2)\,;\\
\end{array}
\]
Thus, $$|pp'|+|qq'|\leq |pq|/s\,.$$ This shows that we only need to
check within distance $|pq|/s$ for possible edges $p'q'$. In P2P
applications, a newly joined peer only need to check its neighboring
nodes, to know whether it should build a direct link to other nodes
or not.}

To show that the algorithm eventually outputs a good spanner, we
first show the connection of $G$ with the notion
\emph{well-separated pair decomposition}.

\begin{definition}[Well-separated pair]
Let $s>0$ be a constant, and a pair of sets of points $A$, $B$ is
$s$-separated, if $d(A,
B)\geq s\cdot \max(\diam(A), \diam(B))$, where $\diam(A)$ is the
diameter of the point set $A$, $\diam(A)=\max\limits_{p, q \in A} |pq|$, and $d(A, B) = \min\limits_{p \in A,
q \in B} |pq|$.
\end{definition}

\begin{definition}[Well-separated pair decomposition]
Let $s>0$ be a constant, and $\P$ be a point set. An $s$-well-separated
pair decomposition (WSPD) of $\P$ is a set of
pairs $\W = \{(A_1, B_1),\ldots, (A_m, B_m) \}$, s.t.
\benum
\item{$A_i, B_i \subseteq P$, and the pair sets $A_i$ and $B_i$ are $s$-separated for every $i$.}
\item For any two points $p, q \in \P$, there is at least one pair $(A_i, B_i)$ such that $p\in A_i$ and $q\in B_i$.
\eenum
Here $m$ is called the size of the WSPD.
\end{definition}

In our construction of $G$, it is not so hard to see that a well-separated pair decomposition is
actually implied.
\begin{theorem}\label{thm:spanner-WSPD}
From the uncoordinated construction of the graph $G$, we can build
the following $s$-well-separated pair decomposition $\W$: for each edge $pq$ in $G$, include in
$\W$ the pair $(B_r(p), B_r(q))$,
with $r=|pq|/(2s+2)$. The size of the WSPD is the number of edges in
$G$.
\end{theorem}
\begin{proof}
First each pair $(B_r(p), B_r(q))$ is an $s$-well-separated pair.
Obviously, $d(B_r(p), B_r(q)) \geq |pq|-2r$, and $\diam(B_r(p)),
\diam(B_r(q)) \leq 2r$. One can then verify that $d(B_r(p), B_r(q))
\geq s\cdot \max(\diam(B_r(p)), \diam(B_r(q)))$.

We now show that any point $p, q$ is included in one well-separated
pair. If the edge $pq$ is in the graph the claim is true obviously.
Otherwise, there is an edge $p'q'$ in $G$ such that $|pp'|\leq
|p'q'|/(2s+2)$, $|qq'|\leq |p'q'|/(2s+2)$, by
Lemma~\ref{lem:separation-lemma}. This means that $p \in B_{r'}(p')$
and $q \in B_{r'}(q')$ with $r'=|p'q'|/(2s+2)$. This finishes the
proof.
\end{proof}

\myparagraph{A greedy algorithm for well-separated pair decomposition.}
The above theorem shows the connection of the uncoordinated graph $G$
with a WSPD $\W$. In fact, the way to compute the WSPD $\W$ via the
construction of $G$ is equivalent to the following algorithm that
computes an $s$-WSPD, in a \emph{greedy} fashion, with $s>1$.
\benum
\item{Choose an arbitrary pair $(p, q)$, not yet covered by existing well-separated pairs in $\W$.}
\item{Include the pair of point sets $B_r(p)$ and $B_r(q)$ in the WSPD $\W$, with $r = |pq|/(2+2s)$.}
\item{Label every point pair $(p_i, q_i)$ with $p_i \in B_r(p)$ and $q_i \in B_r(q)$ as being covered.}
\item Repeat the above steps until every pair of points is covered.
\eenum

With the $s$-WSPD $\W$, the uncoordinate construction of the graph
$G$ is in fact by taking an edge from each and every well-separated
pair in $\W$ --- the simple rule in Lemma~\ref{lem:separation-lemma}
prevented two edges from the same well-separated pair in $\W$ to be
constructed. It is already known that for any well-separated pair
decomposition, if one edge is taken from each well-separated pair,
then the edges will become a spanner on the original point
set~\cite{ams-rdags-94,admss-esstl-95,ns-asfeg-00,levcopoulos02improved}.
For our specific greedy $s$-WSPD, we are able to get a slightly better stretch as shown in the theorem below.
\begin{theorem}\label{thm:stretch}
Graph $G$ constructed from the greedy $s$-WSPD is a spanner with stretch factor $(s+1)/(s-1)$.
\end{theorem}
\begin{proof}
Denote by $\pi(p, q)$ the shortest path length between $p, q$ in the
graph $G$. We show that $\pi(p, q)\leq \beta \cdot |pq|$ for any $p,
q\in P$, with $\beta=(s+1)/(s-1)$. We prove this claim by induction
on the distance between two points $p, q$. Take $p, q$ as the
closest pair of $P$. Then any $s$-WSPD will have to use a singleton
pair $({p}, {q})$ to cover the pair $(p, q)$ if $s> 1$. If
otherwise, say $(P, Q)$ is an $s$-well-separated pair that covers
$(p, q)$, and $|P|>1$. Then $\diam(P)> |pq|$, and $d(P, Q)=|pq|$.
This contradicts with the fact that $d(P, Q)\geq s\cdot \diam(P)$.
Therefore the edge $pq$ is included in $G$ for sure and $\pi(p,
q)=|pq|$.

Now suppose that for all pairs of nodes $x, y$ with Euclidean
distance $|xy|\leq \ell$, we have $\pi(x, y)\leq \beta \cdot |xy|$.
Now we consider the pair of nodes $p, q$ with the smallest distance
(among all remaining pairs) that is still greater than $\ell$. $(p,
q)$ is covered by an $s$-well-separated pair $(P, Q) \in \W$, where
$P=B_r(p')$ and $Q=B_r(q')$ with $r=|p'q'|/(2s+2)$ and $p'q'$ an
edge in $G$. Now we argue that $|pp'|\leq \ell$. If otherwise,
$|pq|\geq d(P, Q)\geq s\cdot \diam(P) \geq s\cdot |pp'|> |pp'|$. So
we should have selected the pair $(p, p')$ instead of $(p, q)$.
Similarly, $|qq'|\leq \ell$. Thus by induction hypothesis $\pi(p,
p')\leq \beta \cdot |pp'|$, $\pi(q, q')\leq \beta \cdot |qq'|$. By
triangle inequality, we have $\pi(p, q)\leq \pi(p, p')+|p'q'|+\pi(q,
q') \leq \beta \cdot (|pp'|+|qq'|)+|p'q'|\leq 2\beta \cdot r +
|p'q'|$. On the other hand, we know by triangle inequality that
$|pq|\geq |p'q'|-2r=\frac{s}{s+1} \cdot |p'q'|$. Combining
everything we get that $\pi(p, q)\leq (\frac{\beta}{s+1}+1)\cdot
\frac{s+1}{s} \cdot |pq|=\beta \cdot |pq|$, with $\beta=(s+1)/(s-1)$. This
finishes the proof.
\end{proof}
To make the stretch factor as $1+\eps$, we just
take $s=1+2/\eps$ in our spanner construction. We also want to show
that the spanner is sparse and has some other nice properties useful
for our applications. For that we will first show that the greedy
WSPD algorithm will output a linear number of well-separated pairs.

\begin{figure*}
\centering \epsfig{file=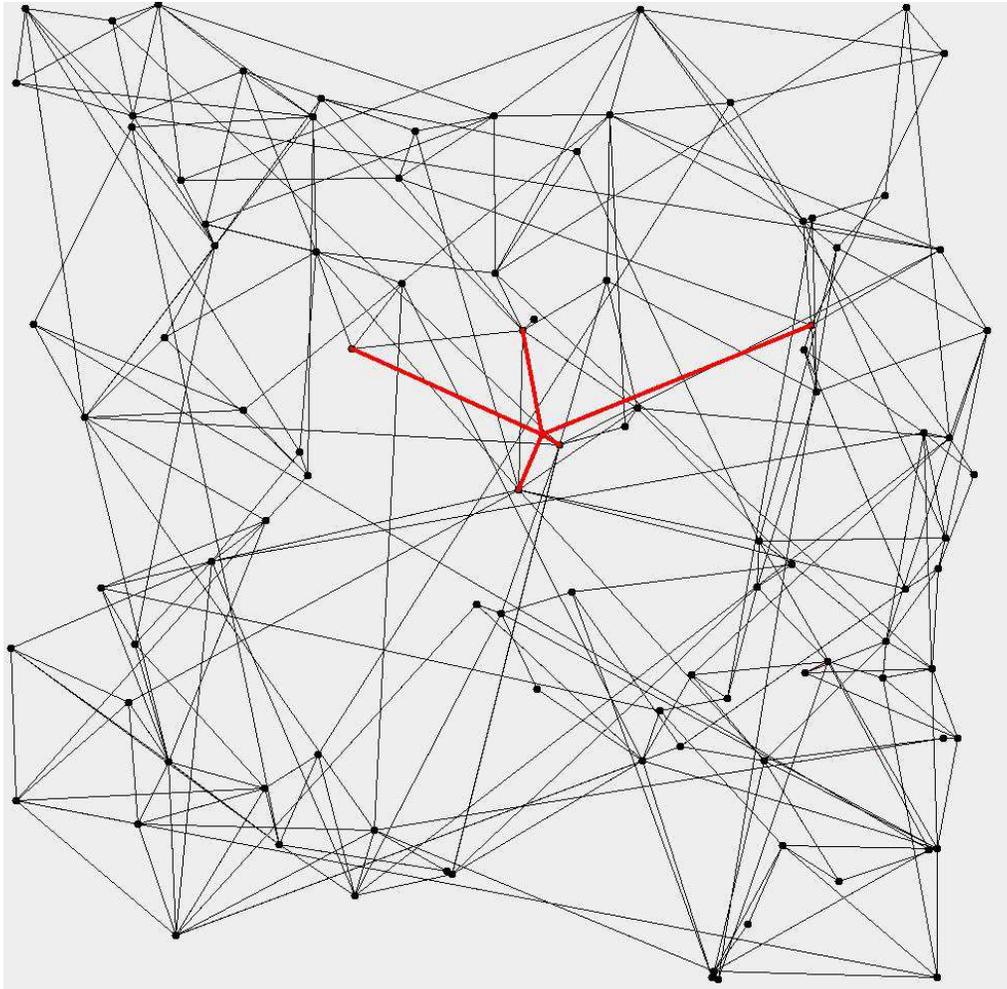, width=0.8\textwidth}
\vspace*{-4mm}\caption{An example of the uncoordinated spanner for 100 points with
aspect ratio $\alpha = 223$, the average degree is 6.5, and the
stretch is 3.4.} \label{fig:spanner} \vspace{-4mm}
\end{figure*}

\section{The uncoordinated spanner has linear size}

To show that the WSPD by the greedy algorithm has a linear number of
pairs, we actually show the connection of this WSPD with a specific
WSPD constructed by the deformable spanner~\cite{gao06deformable},
in the way that at most a constant number of pairs in $\W$ is mapped
to each well-separated pair constructed by the deformable spanner.
To be consistent, in the following description, the greedy WSPD is
denoted by $\W$ and the WSPD constructed by the deformable spanner
is denoted by $\hat{\W}$.

\subsection{Deformable spanner and WSPD}
In this section, we review the basic definition of the deformable
spanner and some related properties, which will be used in our own
algorithm analysis in the next subsection.

Given a set of points $\P$ in the plane, a set of \emph{discrete
centers} with radius $r$ is defined to be the maximal set $S
\subseteq \P$ that satisfies the \emph{covering} property and the
\emph{separation} property: any point $p \in \P$ is within distance
$r$ to some point $p' \in S$; and every two points in $S$ are of
distance at least $r$ away from each other. In other words, all the
points in $\P$ can be covered by balls with radius $r$, whose centers
are exactly those points in the discrete center set $S$. And these
balls do not cover other discrete centers.

We now define a hierarchy of discrete centers in a recursive way.
$S_0$ is the original point set $\P$. $S_i$ is the discrete center
set of $S_{i-1}$ with radius $2^i$. Without loss of generality we
assume that the closest pair has distance $1$ (as we can scale the
point set and do not change the combinatorial structure of the
discrete center hierarchy). Thus the number of levels of the
discrete center hierarchy is $\lg\alpha$, where $\alpha$ is the
aspect ratio of the point set $\P$, defined as the ratio of the
maximum pairwise distance to the minimum pairwise distance, that is,
$\alpha = \max\limits_{u,v\in \P} |uv|/\min\limits_{u,v \in
\P}|uv|$. Since a point $p$ may stay in multiple consecutive levels
and correspond to multiple nodes in the discrete center hierarchy,
we denote by $p^{(i)}$ the existence of $p$ at level $i$. For each
point $p^{(i-1)} \in S_{i-1}$ on level $i-1$, it is within distance
$2^i$ from at least one other point on level $i$. Thus we assign to
$p^{(i-1)}$ a \emph{parent} $q^{(i)}$ in $S_i$ such that
$|p^{(i-1)}q^{(i)}| \leq 2^i$. When there are multiple points in
$S_i$ that cover $p^{(i-1)}$, we choose one as its parent
arbitrarily. We denote by $P(p^{(i-1)})$ the parent of $p^{(i-1)}$
on level $i$. We denote by $P^{(i)}(p)=P(P^{(i-1)}(p))$ the
\emph{ancestor} of $p$ at level $i$.

The deformable spanner is based on the hierarchy, with all edges
between two points $u$ and $v$ in $S_i$ if $|uv| \leq c\cdot 2^i$,
where $c$ is a constant equal to $4+16/\eps$.

We remark that Krauthgamer and Lee~\cite{982913} independently
proposed a very similar hierarchical structure for proximity search
in metrics with doubling dimension~\cite{gupta03bounded}.

Now we will restate some important properties of the deformable
spanner that will be useful in our algorithm analysis.

\begin{lemma}[Packing Lemma~\cite{gao06deformable}]\label{lem:packing}
In a point set $S \subseteq R^d$ , if every two points are at least
distance $r$ away from each other, then there can be at most
$(2R/r+1)^d$ points in $S$ within any ball with radius $R$.
\end{lemma}

\begin{lemma}[Deformable spanner properties~\cite{gao06deformable}]\label{lem:defspanner}
For a set of $n$ points in $R^d$ with aspect ratio $\alpha$, \benum
\item For any point $p \in S_0$, its ancestor $P^{(i)}(p) \in S_i$ is of
distance at most $2^{i+1}$ away from $p$.
\item Any point $p\in S_i$ has at most $(1+2c)^d -1$ edges with other points of $S_i$.
\item The deformable spanner $\hat{G}$ is a $(1+\eps)$-spanner with $O(n/\eps^d)$ edges.
\item $\hat{G}$ has total weight $O(\MST\cdot \lg \alpha/\e ^{d+1} )$, where $\MST$ is the weight of the minimal spanning tree of the point set $S$.
\eenum
\end{lemma}

As shown in~\cite{gao06deformable}, the deformable spanner implies a
well-separated pair decomposition $\hat{\W}$ by taking all the
`cousin pairs'. Specifically, for a node $p^{(i)}$ on level $i$, we
denote by $P_i$ the collection of points that are decedents of
$p^{(i)}$ (including $p^{(i)}$ itself).
Now we take the pair $(P_i, Q_i)$, the sets of decedents of a
\emph{cousin pair} $p^{(i)}$ and $q^{(i)}$, i.e., $p^{(i)}$ and
$q^{(i)}$ are \emph{not} neighbors in level $i$ but their parents
are neighbors in level $i+1$. This collection of pairs constitutes a
$\frac{4}{\eps}$-well-separated pair decomposition. The size of
$\hat{\W}$ is bounded by the number of cousin pairs and is shown to
be in the order of $O(n/\eps^d)$.

\subsection{Greedy well-separated pair decomposition has linear size}

With the WSPD $\hat{\W}$ constructed by the deformable spanner, we
now prove that the greedy WSPD $\W$ has linear size as well. The
basic idea is to map the pairs in $\W$ to the pairs in $\hat{\W}$
and show that at most a constant number of pairs in $\W$ map to the
same pair in $\hat{\W}$.

\begin{theorem}\label{thm:greedy-WSPD}
The greedy $s$-WSPD $\W$ has size $O(ns^d)$.
\end{theorem}

\begin{proof}
Suppose that we have constructed a deformable spanner $DS$ with $c =
4(s+1)$ and obtained an $s$-well-separated pair decomposition (WSPD)
of it, call it $\hat{\W}$, where $s=c/4-1$. The size of $\hat{\W}$
is $O(ns^d)$. Now we
will construct a map that takes each pair in $\W$ and map it to a
pair in $\hat{\W}$.

Each pair $(P, Q)$ in $\W$ is created by considering the points
inside the balls $B_r(p), B_r(q)$ with radius $r = |pq|/(2+2s)$
around $p, q$. Now we consider the ancestors of $p, q$ in the
spanner $DS$ respectively. There is a unique level $i$ such that the
ancestor $u_i=P^{(i)}(p)$ and $v_i=P^{(i)}(q)$ do not have a spanner
edge in between but the ancestor $u_{i+1}=P^{(i+1)}(p)$ and
$v_{i+1}=P^{(i+1)}(q)$ have an edge in between. The pair $u_i$,
$v_i$ is a cousin pair by definition and thus their decedents
correspond to an $s$-well-separated pair $(P_i, Q_i) $ in $\hat{\W}$.
We say that the pair $(B_r(p), B_r(q))\in \W$ maps to the descendant
pair $(P_i, Q_i)\in \hat{\W}$.

By the discrete center hierarchy (Lemma~\ref{lem:defspanner}) and that
$u_i, v_i$ do not have an edge in the spanner, i.e., $|u_iv_i|>c\cdot 2^i$,
we show that, $$|pq|\geq |u_iv_i|-|pu_i|-|qv_i|\geq |u_iv_i|-2\cdot
2^{i+1} \geq (c-4)\cdot 2^i.$$ Also, since $u_{i+1}, v_{i+1}$ have an edge in the spanner,
$|u_{i+1}v_{i+1}|\leq c\cdot 2^{i+1}$, $$|pq|\leq
|pu_{i+1}|+|u_{i+1}v_{i+1}|+|qv_{i+1}|\leq 2\cdot 2^{i+2}+c\cdot
2^{i+1} =2(c+4)\cdot 2^i.$$ Similarly, we have $$c\cdot
2^i<|u_iv_i|\leq |u_iu_{i+1}|+|u_{i+1}v_{i+1}|+|v_iv_{i+1}|\leq
2\cdot 2^{i+1}+c\cdot 2^{i+1}=2(c+2)\cdot 2^{i}.$$ Therefore the
distance between $p$ and $q$ is $c' \cdot |u_iv_i|$, where
$(c-4)/(2c+4) \leq c' \leq (2c+8)/c$.

Now suppose two pairs $(B_{r_1}(p_1), B_{r_1}(q_1))$, $(B_{r_2}(p_2),
B_{r_2}(q_2))$ in $\W$ map to the same pair $u_i$ and $v_i$ by the
above process. Without loss of generality suppose that $p_1, q_1$
are selected before $p_2, q_2$ in our greedy algorithm. Here are some
observations:
\benum
\item{$|p_1q_1| = c_1' \cdot |u_iv_i|$, $|p_2q_2| = c_2' \cdot |u_iv_i|$,
$r_1 =|p_1q_1|/(2+2s)= c_1'\cdot |u_iv_i|/(2+2s)$, $r_2= c_2' \cdot
|u_iv_i|/(2+2s)$, where $(c-4)/(2c+4) \leq c_1', c_2' \leq
(2c+8)/c$, and $r_1$, $r_2$ are the radii of the balls for the two
pairs respectively.}
\item{The reason that $(p_2, q_2)$ can be selected in our greedy algorithm
is that at least one of $p_2$ or $q_2$ is outside the balls $B(p_1),
B(q_1)$, by Lemma~\ref{lem:separation-lemma}. This says that at
least one of $p_2$ or $q_2$ is of distance $r_1$ away from $p_1,
q_1$. }
\eenum

Now we look at all the pairs $(p_\ell, q_\ell)$ that map to the same
ancestor pair $(u_i, v_i)$. The pairs are ordered in the same order
as they are constructed, i.e., $p_1, q_1$ is the first pair selected
in the greedy WSPD algorithm. Suppose $r_{min}$ is the minimum among
all radius $r_i$. $r_{min} = |pq|_{min}/(2+s)\geq (c-4)/(2c+4)\cdot
c\cdot 2^i/(2+2s) = s/(2s+3)\cdot 2^{i+1}$. We group these pairs in
the following way. The first group $H_1$ contains $(p_1, q_1)$ and
all the pairs $(p_\ell, q_\ell)$ that have $p_\ell$ within distance
$r_{min}/2$ from $p_1$. We say that $(p_1, q_1)$ is the
representative pair in $H_1$ and the other pairs in $H_1$ are
\emph{close} to the pair $(p_1, q_1)$. The second group $H_2$
contains, among all remaining pairs, the pair that was selected in
the greedy algorithm the earliest, and all the pairs that are close
to it. We repeat this process to group all the pairs into $k$
groups, $H_1, H_2, \cdots, H_k$. For all the pairs in each group
$H_j$, we have one representative pair, denoted by $(p_j, q_j)$ and
the rest of the pairs in this group are close to it.

We first bound the number of pairs belonging to each group by a
constant with a packing argument. With our group criteria and the
above observations, all $p_\ell$ in the group $H_j$ are within
radius $r_{min}$ from each other. This means that the
$q_\ell$'s must be far away --- the $q_\ell$'s must be at least
distance $r_{min}$ away from each other, by
Lemma~\ref{lem:separation-lemma}. On the other hand, all the
$q_\ell$'s are descendant of the node $v_i$, so $|v_i q_\ell|\leq
2^{i+1}$ by Theorem~\ref{lem:defspanner}. That is, all the
$q_\ell$'s are within a ball of radius $2^{i+1}$ centered at $v_i$.
By the packing Lemma~\ref{lem:packing}, the number of such
$q_\ell$'s is at most $(2\cdot 2^{i+1}/r_{min}+1)^d\leq (2\cdot
2^{i+1} (2s+3)/(s\cdot 2^{i+1})+1)^d = (5 + 6/s)^d$. This is also
the bound on the number of pairs inside each group.

Now we bound the number of different groups, i.e., the value $k$.
For the representative pairs of the $k$ groups, $(p_1, q_1), (p_2,
q_2), \cdots, (p_k, q_k)$, all the $p_i$'s must be at least distance
$r_{min}/2$ away from each other. Again these $p_i$'s are all
descendant of $u_i$ and thus are within distance $2^{i+1}$ from
$u_i$. By a similar packing argument, the number of such $p_i$'s is
bounded by $(4\cdot 2^{i+1}/r_{min}+1)^d \leq (9 + 12/s)^d$. So the
total number of pairs mapped to the same ancestor pair in $\hat{\W}$
will be at most $(5 + 6/s)^d \cdot (9 + 12/s)^d =(O(1+1/s))^d$. Thus
the total number of pairs in $\W$ is at most $O(ns^d)$. This finishes
the proof.
\end{proof}

\section{Size, degree, weight and diameter of the uncoordinated spanner}

With the result that the greedy WSPD has linear size in the previous
section and the connection of the greedy WSPD with the uncoordinated
spanner construction in Section~\ref{sec:spanner}, we are able to obtain the following theorems. The proofs use various of packing arguments.

\begin{theorem}\label{thm:spanner-size}
The uncoordinated spanner $G$ with parameter $s$ is a spanner with
stretch factor $(s+1)/(s-1)$ and has $O(ns^d)$ number of edges.
\end{theorem}
\begin{proof}
The number of edges in the spanner is the same as the size of the
greedy WSPD $\W$ with the same parameter $s$ constructed by
selecting the same set of edges in the same order.
\end{proof}

\begin{figure*}
\centering \epsfig{file=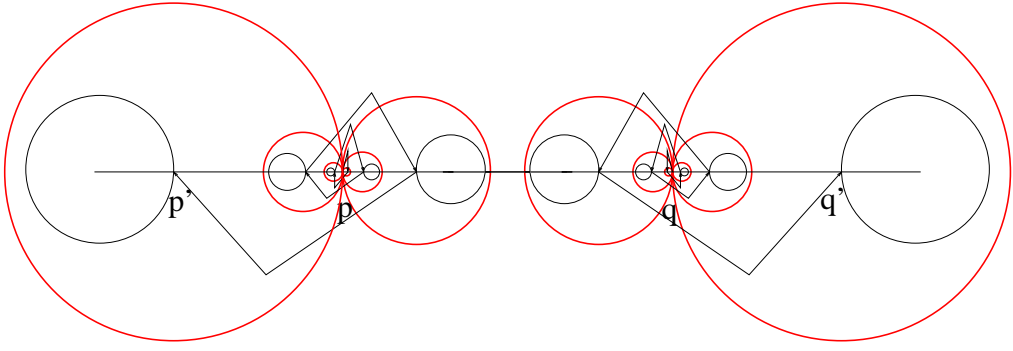, width=0.8\textwidth}
\vspace*{-4mm}\caption{An example shows that we may achieve the
tight bound $2|pq|^{1/(1+\lg s)}$ on the hop distance between  point
$p$ and $q$.} \label{fig:hopbound} \vspace{-4mm}
\end{figure*}

\begin{theorem}\label{thm:spanner-degree}
$G$ has a maximal degree of $O(\lg \alpha
\cdot s^d)$ and average degree $O(s^d)$.
\end{theorem}
\begin{proof}
With the same argument as in Theorem~\ref{thm:greedy-WSPD}, each
pair $(p, q)$ built in the uncoordinated spanner maps
to a pair of ancestors $(P^{(i)}(p), P^{(i)}(q))$ in the deformable
spanner that is a cousin pair. Consider all the edges of $p$ in $G$,
$(p, q_\ell)$, that map to the same ancestor pair $(P^{(i)}(p),
P^{(i)}(q))$. By a similar argument, all the $q_\ell$'s must be at
least distance $r_{min}$ away from each other (since all these pairs
have $p$ as the first element in the pair). Thus we have the number
of such edges is bounded by $(5 + 6/s)^d$. The number of cousin
pairs associated with $P^{(i)}(p)$ is at most $5^d$ times the number
of adjacent edges of $P^{(i+1)}(p)$, and is bounded by $5^d \cdot
[(8s+9)^d-1]$ (by Theorem~\ref{lem:defspanner}). Since there are
$\lfloor \lg \alpha \rfloor$ levels, the total number of edges
associated with the node $p$ is at most $\lfloor \lg \alpha \rfloor
\cdot 5^d \cdot [(8s+9)^d-1] \cdot (5 + 6/s)^d $. Then the maximal
degree of the spanner is $O(\lg \alpha \cdot s^d)$.

Since the spanner has total $O(ns^d)$ edges, the average degree is
$O(s^d)$.
\end{proof}
\begin{theorem}\label{thm:spanner-weight}
$G$ has total weight $O(\lg \alpha \cdot \MST
\cdot s^{d+1})$.
\end{theorem}
\begin{proof}
Again we use the mapping of the uncoordinated spanner edges to the
cousin pairs in the deformable spanner $DS$, as in
Theorem~\ref{thm:greedy-WSPD}. We also use the same notation here.
Consider all the edges $(p_\ell, q_\ell)$ that map to the same
ancestor cousin pair $(u_i, v_i)$. We now map them to the edge
between the parents of this cousin pair, i.e., edge
$u_{i+1}v_{i+1}$ in $DS$. The pair $(u_{i+1}, v_{i+1})$ has at most
$5^{2d}$ number of cousin pairs. Thus at most $(5 + 6/s)^d \cdot (9
+ 12/s)^d \cdot 5^{2d}=(O(1+1/s))^d$ edges in $G$ are mapped to one
edge in $DS$.

Now we will bound the length of an edge $pq$ in $G$ and the edge
$u_{i+1}v_{i+1}$ in $DS$ it maps to. From the proof of
Theorem~\ref{thm:greedy-WSPD}, we know that $(c-4)\cdot 2^i \leq
|pq|\leq 2(c+4)\cdot 2^i$. In addition, $|u_{i+1}v_{i+1}|\leq 2c
\cdot 2^i$ as $u_{i+1}v_{i+1}$ is an edge in $DS$, and
$|u_{i+1}v_{i+1}| \geq |u_iv_i|-|u_{i+1}u_i|-|v_{i+1}v_i| \geq
c\cdot 2^i -2\cdot 2^{i+1}=(c-4)\cdot 2^i$. Thus, $ (c-4)/(2c)\leq
|pq|/|u_{i+1}v_{i+1}| \leq 2(c+4)/(c-4)$.

We now bound the total weight of the spanner $G$. We group all the
edges by the spanner edge in $DS$ that they map to. Thus we have the
total weight of $G$ is at most $2(c+4)/(c-4) \cdot (O(1+1/s))^d$ the
weight of $DS$. By Theorem~\ref{lem:defspanner}, the weight of $DS$
is at most $O(\lg \alpha \cdot \MST \cdot s^{d+1})$. Thus the weight
of $G$ is at most $O(\lg \alpha \cdot \MST \cdot s^{d+1})$.
\end{proof}
\begin{theorem}\label{thm:spanner-hopdiameter}
For any two points $p$ and $q$ in $G$, there is a path with stretch
$(s+1)/(s-1)$ between $p$ and $q$ with at most $2|pq|^{1/(1+\lg s)}$
hops.
\end{theorem}
\begin{proof}
For any two point $p$ and $q$, they will be covered by a pair set
$(P, Q)$ with respect to edge $p'q'$ so that $p \in P$ and $q \in
Q$. The path $p\leadsto q$ between $p$ and $q$ can be found
recursively by taking the path $p\leadsto p'$, then the edge $p'q'$,
and then the path $q'\leadsto q$. This path found recursively will
have stretch $(s+1)/(s-1)$ according to Theorem~\ref{thm:stretch}.

Obviously $|pp'| \leq r \leq |pq|/(2s)$. Denote by $h(|pq|)$ the hop
count of the path between $p$ and $q$ with stretch $(s+1)/(s-1)$.
Thus we can get the following recurrence $h(|pq|) = h(|pp'|)+1 +
h(|qq'|) \leq h(|pq|/(2s))+ 1 + h(|pq|/(2s))$, that is, $h(x) = 2
h(x/(2s)) + 1$. Solve this recurrence we get $h(|pq|) =
2^{\lg_{2s}|pq|+1} = 2^{\lg |pq|/\lg (2s)+1} = 2|pq|^{1/(1+\lg s)}$.

The analysis is tight as shown by Figure~\ref{fig:hopbound}. In the
example, we assume all the nodes almost lying on the line. The 2
balls with the same radius on the left side(or right side) is the
pair that covers $p$(or $q$), and there is an edge between them(in
order to see clearly, we draw curve line between them). The red bold
balls are the balls that contain $p$(or $q$ on the left side). If we
create these pairs in decreasing order on the radius, then we will
have a path between $p$ and $q$ with exactly $2|pq|^{1/(1+\lg s)}$
hops. And we assume all the rest points are far away, then this is
the only path between $p$ and $q$.
\end{proof}

\section{Spanner construction and routing}\label{sec:spanner-construction}

The spanner construction is so far only defined on points in Euclidean space. Now, we will first extend our spanner results to a more general
metric, metric with constant doubling dimension. The doubling
dimension of a metric space $(X, d)$ is the smallest value $\gamma$
such that each ball of radius $R$ can be covered by at most
$2^\gamma $ balls of radius $R/2$~\cite{gupta03bounded}. Metrics with constant doubling dimension appear in many different settings. For example, it has been discovered that the Internet delay metric has restricted growth rate
and approximately has a constant doubling
dimension~\cite{ng02predicting,plaxton97accessing}. As the main technique used in the spanner analysis is packing argument, the following Theorem is the extension of the spanner in the metric case. The proof is in the Appendix.

\begin{theorem}\label{thm:spanner-doubling-metric}
For $n$ points and a metric space defined on them with constant
doubling dimension $\gamma$, the uncoordinated spanner construction
outputs a spanner $G$ with stretch factor $(s+1)/(s-1)$, has total
weight $O(\gamma ^9 \cdot \lg \alpha \cdot |MST|\cdot
s^{O(\gamma)})$ and has $O(\gamma^4\cdot n\cdot s^{O(\gamma)})$
number of edges. Also it has a maximal degree of $O(\gamma^4 \cdot
\lg\alpha\cdot s^{O(\gamma)})$ and average degree $O(\gamma^4 \cdot
s^{O(\gamma)})$.
\end{theorem}
\begin{proof}
The proof follows almost the same as those in the previous section.
The deformable spanner can be applied for metrics with constant
doubling dimension~\cite{gao06deformable}. Whenever we use a
geometric packing argument, we replace by the property of metrics of
constant doubling dimension. We just need to notice that, with the
proposition from ~\cite{gupta03bounded}, the number of pairs in a
group that might be mapped to the same ancestor is bounded by
$\gamma ^ {\lceil \lg (2R/r)\rceil} = \gamma ^{\lceil \lg (2\cdot
2^{i+1}/(s/(2s+3)\cdot 2^{i+1}))\rceil} = \gamma ^{1+\lceil
\lg(2+3/s)\rceil}< \gamma ^4$, and the number of groups is bounded
by $\gamma ^ {\lceil \lg (2R/(r/2))\rceil} = \gamma ^{\lceil \lg
(4\cdot 2^{i+1}/(s/(2s+3)\cdot 2^{i+1}))\rceil} < \gamma ^5$,
\end{proof}

\myparagraph{Model of computing.} Now we would like to discuss the
computing model as well as the construction cost of an uncoordinated spanner. Each distributed agent checks whether an edge to another agent should be built or not. The order for these operations can be arbitrary. We assume that there is already an oracle that answers near neighbors queries: return the list of nodes within distance $r$ from any node $x$. Such near neighbor oracle may be already available in the applications. For example, in transportation network, if we think each city as an agent, it usually stores the nearby cities and the corresponding transportation road by them. For a metric with doubling dimension, this oracle can be implemented in a number of ways~\cite{plaxton97accessing,982913,hildrum04note}. For example, such functions can be implemented in Tapestry, a P2P system~\cite{hildrum02distributed,Tapestry}.


\myparagraph{Spanner construction and representation.} The spanner
edges are recorded in a distributed fashion so that no node has the
entire picture of the spanner topology. After each edge $pq$ in $G$
is constructed, the peers $p, q$ will inform their neighboring nodes
(those in $B_r(p)$ and $B_r(q)$ with $r=|pq|/(2s+2)$) that such an
edge $pq$ exists so that they will not try to connect to one
another. These neighboring nodes can be obtained through a near neighbor search from both $p$ and $q$. We assume that these messages are delivered immediately so
that when any edge is built the previous constructed edges have been
informed to nodes of relevance. We remark that the
nodes in $B_r(p)$ and $B_r(q)$ will only store the single edge $pq$,
as well as their distance to $p$ (for those in $B_r(p)$) or $q$ (for
those in $B_r(q)$) and the value of $|pq|$, but not all the nodes in
$B_r(p)$ or $B_r(q)$. When a node $x$ considers whether an edge $xy$ should be built, $x$ could simply consult with the edges that it has stored locally. The amount of storage at each node $x$ is
proportional to the number of well-separated pairs that include $x$.
The number of messages for this operation is bounded by $|B_r(p)| +
|B_r(q)|$.
The following theorem shows that the total number of such messages
during the execution of the algorithm is almost linear in $n$. That
is, on average each node sends about $O(s^d \cdot \lg \alpha)$
messages. Also the amount of storage at each node is bounded by
$O(s^d \lg \alpha)$.

\begin{theorem}\label{thm:WSP-per-node}
For the uncoordinated spanner $G$ and the corresponding greedy WSPD
$\W = \{(P_i, Q_i)\}$ with size $m$, each node $x$ is included in at
most $O(s^d \lg \alpha)$ well-separated pairs in $\W$.
\end{theorem}
\begin{proof}
For each point $x$, each well-separated pair that contains $x$ can
be mapped to some ancestor pair in some level in the corresponding
deformable spanner. Let's consider the set pair $(P, Q)$ (with
respect to edge $pq$) that is mapped to the ancestor pair
$(P^{(i)}(p), P^{(i)}(q))=(u_i, v_i)$ at level $i$ with $x \in P$.
Now we map $x$ to the node $u_i$ and we count how many such nodes
$u_i$ on level $i$ that a node $x$ may map to.
The corresponding radius with $(p, q)$ is $r = |pq|/(2+2s) \leq
(1+1/(s+1))2^{i+2}$. In this case, $|xp_i|\leq |pu_i|+|xp|\leq
2^{i+1} + r \leq (3+2/(s+1))2^{i+1}$. According to
Lemma~\ref{lem:defspanner}, different nodes in level $i$ must be at
least $2^i$ away from each other. With a packing argument, there can
be at most $[((2 \cdot (3+2/(s+1))) \cdot 2^{i+1})/2^{i}+1]^d =
(13+8/(s+1))^d$ different such nodes $p_i$ on level $i$ that $x$ may
map to. So the number of well-separated pairs that cover $x$ is at
most
\[
\begin{array}{ll}
&\sum_{i=1}^{\lg\alpha}\sum_{p_i}\big|\mbox{cousin pairs with }p_i\big|\cdot \big|\mbox{pairs in }\W \mbox{ mapping to a cousin pair with }p_i\big|\\
\leq & \lg\alpha \cdot (13+8/(s+1))^d \cdot 5^d [(8s+9)^d-1]\cdot (O(1+1/s)^d)\\
= & O(s^d\lg\alpha).
\end{array}
\]
\end{proof}

\begin{theorem}\label{thm:msg}
For the uncoordinated spanner $G$ and the corresponding greedy WSPD
$\W = \{(P_i, Q_i)\}$ with size $m$, $\sum_{i=1}^m (|P_i|+|Q_i|) =
O(n s^d \cdot \lg \alpha)$. Thus the total messages involved in the spanner construction algorithm is $O(n s^d \cdot \lg \alpha)$.
\end{theorem}
\begin{proof}
This follows immediately from Theorem~\ref{thm:WSP-per-node}, as
$$\sum_{i=1}^m (|P_i|+|Q_i|)\leq \sum_{x} |\mbox{pairs that include
}x|=O(n s^d \cdot \lg \alpha).$$
\end{proof}

\myparagraph{Distributed low-stretch routing on spanner.} Although
the spanner topology is implicitly stored on the nodes with each
node only knows some piece of it, we are actually able to do a
distributed and local routing on the spanner with only information
available at the nodes such that the path discovered has maximum
stretch $(s+1)/(s-1)$. In particular, for any node $p$ who has a
message to send to node $q$, it is guaranteed that $(p, q)$ is
covered by a well-separated pair $(B_r(p'), B_r(q'))$ with $p \in
B_r(p')$ and $q \in B_r(q')$. By the construction algorithm, the
edge $p'q'$, after constructed, is informed to all nodes in
$B_r(p')\cup B_r(q')$, including $p$. Thus $p$ includes in the
packet a partial route with $\{p\rightsquigarrow p', p'q',
q'\rightsquigarrow q\}$. The notation $p\rightsquigarrow p'$ means
that $p$ will need to first find out the low-stretch path from $p$
to the node $p'$ (inductively), from where the edge $p'q'$ can be
taken, such that with another low-stretch path to be found out from
$q'$ to $q$, the message can be delivered to $q$. This way of
routing with partial routing information stored with the packet is
similar to the idea of source routing~\cite{tanenbaum96computer}
except that we do not include the full routing path at the source
node. By the same induction as used in the proof of spanner stretch
(Theorem~\ref{thm:stretch}), the final path is going to have stretch
at most $(s+1)/(s-1)$ and at most $2|pq|^{1/(1+\lg s)}$ hops.

\myparagraph{Support for nearest neighbor search.} The constructed
spanner can be used to look for the nearest peer in the P2P network.
Since we let each point $x$ keep all the edges $(p, q)$ that cover
$x$, among all these $p$'s, one of them must be the nearest neighbor
of $x$. If otherwise, suppose $y$ is the nearest neighbor of $x$,
and $y$ is not one of $p$. But in the WSPD $\W$, $(x, y)$ will
belong to one of the pair set $(P, Q)$, which corresponds to a
spanner edge $(p', q')$ that covers $x$. Then there is a
contradiction, as $|xp'| \leq \diam(P) \leq d(P, Q)/s <  d(P, Q)
\leq |xy|$ implies that $y$ is not the nearest neighbor of $x$.
According to Theorem~\ref{thm:WSP-per-node}, $x$ will belong to at
most $O(s^d\lg\alpha)$ different pair sets. So the nearest neighbor
search can be done in $O(s^d\lg\alpha)$ time, with only the
information stored on $x$.

\myparagraph{Node insertion and deletion.} The uncoordinated spanner
construction supports node insertion and deletion. When a
peer $x$ joins the network, it will check with each other peer
whether or not a nearby edge exists as specified in our greedy
algorithm. When a peer $y$ leaves the network, $p$ will notify the
nodes that are covered by $p$'s edges, i.e., for each edge $pq$, $p$
will notify $q$, and all the nodes within $|pq|/(2s+2)$ from $p$ and
$q$. Then the notified nodes will check and possibly build new edges
to restore the spanner property. In this way, the spanner and the
good properties can be maintained.

\section{Conclusion and future work}

This paper aims to explain the emergence of good spanners from the
behaviors of agents with their own interests. The results can be
immediately applied to the construction of good network overlays by
distributed peers with incomplete information. For our future work
we would like to explore incentive-based overlay
construction~\cite{feigenbaum02distributed}. One problem faced in
the current P2P system design is to reward peers that contribute to
the network maintenance or service quality and punish the peers that
try to take free
rides~\cite{feldman04robust,habib04incentive,habib06service,feldman05overcoming,schosser06incentives,schosser07indirect}.
We would like to extend the results in this paper and come up with a
spanner construction with different quality of service for different
peers to achieve fairness --- those who build more edges should have
a smaller stretch to all other nodes and those who do not build many
edges are punished accordingly by making the distances to others
slightly longer.

\begin{small}
\bibliographystyle{abbrv}
\bibliography{gwspd-full}

\begin{thebibliography}{10}

\bibitem{albers06susanne}
S.~Albers, S.~Eilts, E.~Even-Dar, Y.~Mansour, and L.~Roditty.
\newblock On nash equilibria for a network creation game.
\newblock In {\em SODA '06: Proceedings of the seventeenth annual ACM-SIAM
  symposium on Discrete algorithm}, pages 89--98, New York, NY, USA, 2006. ACM.

\bibitem{admss-esstl-95}
S.~Arya, G.~Das, D.~M. Mount, J.~S. Salowe, and M.~Smid.
\newblock Euclidean spanners: short, thin, and lanky.
\newblock In {\em Proc. 27th ACM Symposium on Theory Computing}, pages
  489--498, 1995.

\bibitem{ams-rdags-94}
S.~Arya, D.~M. Mount, and M.~Smid.
\newblock Randomized and deterministic algorithms for geometric spanners of
  small diameter.
\newblock In {\em Proc. 35th IEEE Symposium on Foundations of Computer
  Science}, pages 703--712, 1994.

\bibitem{as-ecbds-97}
S.~Arya and M.~Smid.
\newblock Efficient construction of a bounded-degree spanner with low weight.
\newblock {\em Algorithmica}, 17:33--54, 1997.

\bibitem{callahan93faster}
Callahan and Kosaraju.
\newblock Faster algorithms for some geometric graph problems in higher
  dimensions.
\newblock In {\em Proc. 4th ACM-SIAM Symposium on Discrete Algorithms}, pages
  291--300, 1993.

\bibitem{c-opann-93}
P.~B. Callahan.
\newblock Optimal parallel all-nearest-neighbors using the well-separated pair
  decomposition.
\newblock In {\em Proc. 34th IEEE Symposium on Foundations of Computer
  Science}, pages 332--340, 1993.

\bibitem{ck-adcpn-95}
P.~B. Callahan and S.~R. Kosaraju.
\newblock Algorithms for dynamic closest-pair and {$n$}-body potential fields.
\newblock In {\em Proc. 6th ACM-SIAM Symposium on Discrete Algorithms}, pages
  263--272, 1995.

\bibitem{ck-dmpsa-95}
P.~B. Callahan and S.~R. Kosaraju.
\newblock A decomposition of multidimensional point sets with applications to
  $k$-nearest-neighbors and $n$-body potential fields.
\newblock {\em J. {ACM}}, 42:67--90, 1995.

\bibitem{chan05hierarchical}
H.~T.-H. Chan, A.~Gupta, B.~M. Maggs, and S.~Zhou.
\newblock On hierarchical routing in doubling metrics.
\newblock In {\em SODA '05: Proceedings of the sixteenth annual ACM-SIAM
  symposium on Discrete algorithms}, pages 762--771, 2005.

\bibitem{small06chan}
T.-H.~H. Chan and A.~Gupta.
\newblock Small hop-diameter sparse spanners for doubling metrics.
\newblock In {\em SODA '06: Proceedings of the seventeenth annual ACM-SIAM
  symposium on Discrete algorithm}, pages 70--78, New York, NY, USA, 2006. ACM.

\bibitem{cdns-nsrgs-95}
B.~Chandra, G.~Das, G.~Narasimhan, and J.~Soares.
\newblock New sparseness results on graph spanners.
\newblock {\em Internat. J. Comput. Geom. Appl.}, 5:125--144, 1995.

\bibitem{chu01enabling}
Y.~Chu, S.~Rao, S.~Seshan, and H.~Zhang.
\newblock Enabling conferencing applications on the internet using an overlay
  muilticast architecture.
\newblock {\em SIGCOMM Comput. Commun. Rev.}, 31(4):55--67, 2001.

\bibitem{corbo05price}
J.~Corbo and D.~Parkes.
\newblock The price of selfish behavior in bilateral network formation.
\newblock In {\em PODC '05: Proceedings of the twenty-fourth annual ACM
  symposium on Principles of distributed computing}, pages 99--107, New York,
  NY, USA, 2005. ACM.

\bibitem{dhn-oss3d-93}
G.~Das, P.~Heffernan, and G.~Narasimhan.
\newblock Optimally sparse spanners in $3$-dimensional {Euclidean} space.
\newblock In {\em Proc. 9th Annu. ACM Sympos. Comput. Geom.}, pages 53--62,
  1993.

\bibitem{dn-facse-97}
G.~Das and G.~Narasimhan.
\newblock A fast algorithm for constructing sparse {Euclidean} spanners.
\newblock {\em Internat. J. Comput. Geom. Appl.}, 7:297--315, 1997.

\bibitem{dns-nwwme-95}
G.~Das, G.~Narasimhan, and J.~Salowe.
\newblock A new way to weigh malnourished {Euclidean} graphs.
\newblock In {\em Proc. 6th ACM-SIAM Sympos. Discrete Algorithms}, pages
  215--222, 1995.

\bibitem{e-sts-00}
D.~Eppstein.
\newblock Spanning trees and spanners.
\newblock In J.-R. Sack and J.~Urrutia, editors, {\em Handbook of Computational
  Geometry}, pages 425--461. Elsevier Science Publishers B.V. North-Holland,
  Amsterdam, 2000.

\bibitem{e-dpssd-02}
J.~Erickson.
\newblock Dense point sets have sparse {D}elaunay triangulations.
\newblock In {\em Proc. 13th ACM-SIAM Symposium on Discrete Algorithms}, pages
  125--134, 2002.

\bibitem{fabrikant03network}
A.~Fabrikant, A.~Luthra, E.~Maneva, C.~H. Papadimitriou, and S.~Shenker.
\newblock On a network creation game.
\newblock In {\em PODC '03: Proceedings of the twenty-second annual symposium
  on Principles of distributed computing}, pages 347--351, 2003.

\bibitem{feigenbaum02distributed}
J.~Feigenbaum and S.~Shenker.
\newblock Distributed algorithmic mechanism design: recent results and future
  directions.
\newblock In {\em DIALM '02: Proceedings of the 6th international workshop on
  Discrete algorithms and methods for mobile computing and communications},
  pages 1--13, New York, NY, USA, 2002. ACM.

\bibitem{feldman05overcoming}
M.~Feldman and J.~Chuang.
\newblock Overcoming free-riding behavior in peer-to-peer systems.
\newblock {\em SIGecom Exch.}, 5(4):41--50, 2005.

\bibitem{feldman04robust}
M.~Feldman, K.~Lai, I.~Stoica, and J.~Chuang.
\newblock Robust incentive techniques for peer-to-peer networks.
\newblock In {\em EC '04: Proceedings of the 5th ACM conference on Electronic
  commerce}, pages 102--111, New York, NY, USA, 2004. ACM.

\bibitem{gao06deformable}
J.~Gao, L.~Guibas, and A.~Nguyen.
\newblock Deformable spanners and their applications.
\newblock {\em Computational Geometry: Theory and Applications}, 35(1-2):2--19,
  2006.

\bibitem{gottlieb08improved}
L.-A. Gottlieb and L.~Roditty.
\newblock Improved algorithms for fully dynamic geometric spanners and
  geometric routing.
\newblock In {\em SODA '08: Proceedings of the nineteenth annual ACM-SIAM
  symposium on Discrete algorithms}, pages 591--600, Philadelphia, PA, USA,
  2008. Society for Industrial and Applied Mathematics.

\bibitem{glns-adogg-02}
J.~Gudmundsson, C.~Levcopoulos, G.~Narasimhan, and M.~Smid.
\newblock Approximate distance oracles for geometric graphs.
\newblock In {\em Proc. 13th ACM-SIAM Symposium on Discrete Algorithms}, pages
  828--837, 2002.

\bibitem{gupta03bounded}
A.~Gupta, R.~Krauthgamer, and J.~R. Lee.
\newblock Bounded geometries, fractals, and low-distortion embeddings.
\newblock In {\em FOCS '03: Proceedings of the 44th Annual IEEE Symposium on
  Foundations of Computer Science}, pages 534--543, 2003.

\bibitem{habib04incentive}
A.~Habib and J.~Chuang.
\newblock Incentive mechanism for peer-to-peer media streaming.
\newblock In {\em Proc. of the 12th IEEE International Workshop on Quality of
  Service (IWQoS'04)}, June 2004.

\bibitem{habib06service}
A.~Habib and J.~Chuang.
\newblock Service differentiated peer selection: An incentive mechanism for
  peer-to-peer media streaming.
\newblock {\em IEEE Transactions on Multimedia}, 8(3):610--621, June 2006.

\bibitem{Har-Peled05fast}
S.~Har-Peled and M.~Mendel.
\newblock Fast construction of nets in low dimensional metrics, and their
  applications.
\newblock In {\em SCG '05: Proceedings of the twenty-first annual symposium on
  Computational geometry}, pages 150--158, New York, NY, USA, 2005. ACM.

\bibitem{hildrum04note}
K.~Hildrum, J.~Kubiatowicz, S.~Ma, and S.~Rao.
\newblock A note on the nearest neighbor in growth-restricted metrics.
\newblock In {\em SODA '04: Proceedings of the fifteenth annual ACM-SIAM
  symposium on Discrete algorithms}, pages 560--561, Philadelphia, PA, USA,
  2004. Society for Industrial and Applied Mathematics.

\bibitem{hildrum02distributed}
K.~Hildrum, J.~D. Kubiatowicz, S.~Rao, and B.~Y. Zhao.
\newblock Distributed object location in a dynamic network.
\newblock In {\em SPAA '02: Proceedings of the fourteenth annual ACM symposium
  on Parallel algorithms and architectures}, pages 41--52, New York, NY, USA,
  2002. ACM.

\bibitem{jansen07stability}
T.~Jansen and M.~Theile.
\newblock Stability in the self-organized evolution of networks.
\newblock In {\em GECCO '07: Proceedings of the 9th annual conference on
  Genetic and evolutionary computation}, pages 931--938, New York, NY, USA,
  2007. ACM.

\bibitem{982913}
R.~Krauthgamer and J.~R. Lee.
\newblock Navigating nets: simple algorithms for proximity search.
\newblock In {\em Proceedings of the fifteenth annual ACM-SIAM symposium on
  Discrete algorithms}, pages 798--807, 2004.

\bibitem{kwon02topology}
M.~Kwon and S.~Fahmy.
\newblock Topology-aware overlay networks for group communication.
\newblock In {\em NOSSDAV '02: Proceedings of the 12th international workshop
  on Network and operating systems support for digital audio and video}, pages
  127--136, New York, NY, USA, 2002. ACM.

\bibitem{levcopoulos02improved}
C.~Levcopoulos, G.~Narasimhan, and M.~H.~M. Smid.
\newblock Improved algorithms for constructing fault-tolerant spanners.
\newblock {\em Algorithmica}, 32(1):144--156, 2002.

\bibitem{lua05survey}
K.~Lua, J.~Crowcroft, M.~Pias, R.~Sharma, and S.~Lim.
\newblock A survey and comparison of peer-to-peer overlay network schemes.
\newblock {\em Communications Surveys \& Tutorials, IEEE}, pages 72--93, 2005.

\bibitem{moscibroda06topologies}
T.~Moscibroda, S.~Schmid, and R.~Wattenhofer.
\newblock On the topologies formed by selfish peers.
\newblock In {\em PODC '06: Proceedings of the twenty-fifth annual ACM
  symposium on Principles of distributed computing}, pages 133--142, New York,
  NY, USA, 2006. ACM.

\bibitem{ns-asfeg-00}
G.~Narasimhan and M.~Smid.
\newblock Approximating the stretch factor of {Euclidean} graphs.
\newblock {\em SIAM J. Comput.}, 30:978--989, 2000.

\bibitem{geometric07narasimhan}
G.~Narasimhan and M.~Smid.
\newblock {\em Geometric Spanner Networks}.
\newblock Cambridge University Press, 2007.

\bibitem{ng02predicting}
E.~Ng and H.~Zhang.
\newblock Predicting {Internet} network distance with coordinates-based
  approaches.
\newblock In {\em Proc. IEEE INFOCOM}, pages 170--179, 2002.

\bibitem{peleg00}
D.~Peleg.
\newblock {\em Distributed Computing: A Locality Sensitive Approach}.
\newblock Monographs on Discrete Mathematics and Applications. SIAM, 2000.

\bibitem{plaxton97accessing}
C.~G. Plaxton, R.~Rajaraman, and A.~W. Richa.
\newblock Accessing nearby copies of replicated objects in a distributed
  environment.
\newblock In {\em Proc. {ACM} Symposium on Parallel Algorithms and
  Architectures}, pages 311--320, 1997.

\bibitem{ratnasamy02topologically}
S.~Ratnasamy, M.~Handley, R.~Karp, and S.~Shenker.
\newblock Topologically-aware overlay construction and server selection.
\newblock In {\em Proceedings of the 21th Annual Joint Conference of the IEEE
  Computer and Communications Societies (INFOCOM'05)}, volume~3, pages
  1190--1199, 2002.

\bibitem{roditty07fully}
L.~Roditty.
\newblock Fully dynamic geometric spanners.
\newblock In {\em SCG '07: Proceedings of the twenty-third annual symposium on
  Computational geometry}, pages 373--380, New York, NY, USA, 2007. ACM.

\bibitem{schosser06incentives}
S.~Schosser, K.~B\"{o}hm, R.~Schmidt, and B.~Vogt.
\newblock Incentives engineering for structured p2p systems - a feasibility
  demonstration using economic experiments.
\newblock In {\em EC '06: Proceedings of the 7th ACM conference on Electronic
  commerce}, pages 280--289, New York, NY, USA, 2006. ACM.

\bibitem{schosser07indirect}
S.~Schosser, K.~B\"{o}hm, and B.~Vogt.
\newblock Indirect partner interaction in peer-to-peer networks: stimulating
  cooperation by means of structure.
\newblock In {\em EC '07: Proceedings of the 8th ACM conference on Electronic
  commerce}, pages 124--133, New York, NY, USA, 2007. ACM.

\bibitem{tanenbaum96computer}
A.~S. Tanenbaum.
\newblock {\em Computer networks (3rd ed.)}.
\newblock Prentice-Hall, Inc., Upper Saddle River, NJ, USA, 1996.

\bibitem{wang05network}
W.~Wang, C.~Jin, and S.~Jamin.
\newblock Network overlay construction under limited end-to-end reachability.
\newblock In {\em Proceedings of the 24th Annual Joint Conference of the IEEE
  Computer and Communications Societies (INFOCOM'05)}, volume~3, pages
  2124--2134, March 2005.

\bibitem{zhang08distributed}
X.~Zhang, Z.~Li, and Y.~Wang.
\newblock A distributed topology-aware overlays construction algorithm.
\newblock In {\em MG '08: Proceedings of the 15th ACM Mardi Gras conference},
  pages 1--6, New York, NY, USA, 2008. ACM.

\bibitem{Tapestry}
B.~Y. Zhao, J.~D. Kubiatowicz, and A.~D. Joseph.
\newblock Tapestry: An infrastructure for fault-tolerant wide-area location
  and.
\newblock Technical report, Berkeley, CA, USA, 2001.

\end{thebibliography}
\end{small}

\end{document}